\documentclass[sigconf]{acmart}
\usepackage{balance}
\usepackage{paralist}
\usepackage{multirow}
\usepackage{multicol}
\usepackage{anyfontsize}
\usepackage{macros}

\setcopyright{acmcopyright}

\newtoggle{withappendix}
\toggletrue{withappendix}

\makeatletter
\algrenewcommand\ALG@beginalgorithmic{\small}
\makeatother
\algrenewcommand\algorithmicindent{0.3cm}

\begin{document}
	
\copyrightyear{2019} 
\acmYear{2019} 
\acmConference[CIKM '19]{The 28th ACM International Conference on Information and Knowledge Management}{November 3--7, 2019}{Beijing, China}
\acmBooktitle{The 28th ACM International Conference on Information and Knowledge Management (CIKM '19), November 3--7, 2019, Beijing, China}
\acmPrice{15.00}
\acmDOI{10.1145/3357384.3358147}
\acmISBN{978-1-4503-6976-3/19/11}

\fancyhead{}

\author{Pan Hu}
\affiliation{%
  \institution{University of Oxford}}

\author{Jacopo Urbani}
\affiliation{%
  \institution{Vrije Universiteit Amsterdam}}

\author{Boris Motik}
\affiliation{%
  \institution{University of Oxford}}

\author{Ian Horrocks}
\affiliation{%
  \institution{University of Oxford}}

\title{Datalog Reasoning over Compressed RDF Knowledge Bases}

\begin{abstract}
\emph{Materialisation} is often used in RDF systems as a preprocessing step to
derive all facts implied by given RDF triples and rules. Although widely used,
materialisation considers all possible rule applications and can use a lot of
memory for storing the derived facts, which can hinder performance. We present
a novel materialisation technique that compresses the RDF triples so that the
rules can sometimes be applied to multiple facts at once, and the derived facts
can be represented using structure sharing. Our technique can thus require less
space, as well as skip certain rule applications. Our experiments show that our
technique can be very effective: when the rules are relatively simple, our
system is both faster and requires less memory than prominent state-of-the-art
RDF systems.

\end{abstract}

\ccsdesc[500]{Information systems~Data management systems}

\maketitle

\section{Introduction}\label{sec:introduction}

Datalog \cite{abiteboul95foundation} is a prominent knowledge representation
language that can describe an application domain declaratively using if--then
rules. Datalog applications typically require answering queries over facts
derived from knowledge bases (KBs) encoded on the Web using the RDF
\cite{rdf11-concepts} data model. Modern datalog-based RDF systems, such as
graphDB \cite{DBLP:journals/semweb/BishopKOPTV11}, Oracle's RDF Database
\cite{wu2008implementing}, VLog \cite{DBLP:conf/aaai/UrbaniJK16}, and RDFox
\cite{mnpho14parallel-materialisation-RDFox}, derive and store all implied
facts in a preprocessing step. This style of reasoning is commonly called
\emph{materialisation} and is widely used since it enables efficient query
answering. Despite its popularity, however, such an approach exhibits two main
drawbacks. First, deriving all implied facts requires considering all possible
inferences (i.e., applications of the rules to facts). The number of inferences
can be very large (i.e., worst-case exponential in the number of variables in
the rules), so materialisation can take a long time when the KB is large.
Second, the rules can derive a large number of facts, which can impose
significant memory requirements on datalog systems.

In this paper, we present a novel technique for materialising datalog rules
over RDF datasets, which aims to address both of these problems. We observed
that the facts in KBs often exhibit a degree of regularity. For example, facts
denoting similar items in a product catalog of an e-commerce application are
likely to be similar. This often leads to regular rule applications: rules are
usually applied to similar facts in similar ways, and they produce similar
conclusions. We exploit this regularity to address both sources of inefficiency
outlined above. To reduce the memory usage, we represent the derived facts
using \emph{structure sharing}---that is, we store the common parts of facts
only once. This, in turn, allows us to apply certain rules to several facts at
once and thus skip certain rule applications.

We borrow ideas from columnar databases \cite{DBLP:journals/debu/IdreosGNMMK12}
to represent facts. For example, to represent RDF triples
${\triple{a_1}{P}{b_1}, \dots, \triple{a_n}{P}{b_n}}$, we sort the triples and
represent them using just one \emph{meta-fact} ${P(\mc a, \mc b)}$, where
\emph{meta-constants} $\mc a$ and $\mc b$ are sorted vectors of constants
${a_1, \dots, a_n}$ and ${b_1, \dots, b_n}$, respectively. Columnar databases
can efficiently compute joins over such a representation
\cite{DBLP:conf/sigmod/AbadiMF06,DBLP:journals/pvldb/LambFVTVDB12,Abadi:2008:QEC:1467436,DBLP:conf/vldb/ManegoldBKN04}.
However, these techniques address only a part of the problem since, during
materialisation, join evaluation is constantly interleaved with database
updates and elimination of duplicate facts (which is needed for termination).
The VLog system was among the first to use a columnar representation of facts,
and it optimises application of rules with one body atom. However, on complex
rules, VLog computes joins and represents their consequences as usual.

We take these ideas one step further and present algorithms that (i)~can handle
arbitrary joins in rule bodies, (ii)~aim to represent the derived facts
compactly, and (iii)~can efficiently eliminate duplicate facts. We have
implemented our techniques in a new system called CompMat and have empirically
compared it with VLog and RDFox on several well-known benchmarks. Our
experiments show that our technique can sometimes represent the materialisation
by an order of magnitude more compactly than as a `flat' list of facts, thus
allowing our system to handle larger KBs without additional memory. Moreover,
our prototype could often compute the materialisation more quickly than
existing highly-optimised RDF systems.

\section{Preliminaries}\label{sec:preliminaries}

Datalog knowledge bases are constructed using \emph{constants},
\emph{variables}, and \emph{predicates}, where each predicate is associated
with a nonnegative integer called \emph{arity}. A \emph{term} is a constant or
a variable. An \emph{atom} has the form ${P(t_1,\dots,t_n)}$, where $P$ is an
$n$-ary predicate and each $t_i$ is a term. A \emph{fact} is a variable-free
atom, and a \emph{dataset} is a finite set of facts. A \emph{(datalog) rule}
$r$ has the form ${B_1 \wedge \dots \wedge B_n \rightarrow H}$ where ${n \geq
0}$, $H$ is a \emph{head atom}, $B_i$ are \emph{body atoms}, and each variable
in $H$ occurs in some $B_i$. A \emph{(datalog) program} $\Pi$ is a finite set
of rules.

A \emph{substitution} $\sigma$ is a mapping of variables to constants, and
$\dom{\sigma}$ is the domain of $\sigma$. For $\alpha$ a logical expression,
$\alpha\sigma$ is the result of replacing each occurrence in $\alpha$ of a
variable ${x \in \dom{\sigma}}$ with $\sigma(x)$. For $I$ a dataset and ${r =
B_1 \wedge \dots \wedge B_n \rightarrow H}$ a rule, the result of applying $r$
to $I$ is given by ${\apply{r}{I} = \{ H\sigma \mid \{ B_1\sigma, \dots,
B_n\sigma \} \subseteq I \}}$; analogously, for $\Pi$ a program,
${\apply{\Pi}{I} = \bigcup\textstyle_{r \in \Pi} \apply{r}{I}}$. Given a
dataset $E$ of \emph{explicitly given facts}, let ${I_0 = E}$, and for ${i \geq
1}$ let ${I_i = I_{i-1} \cup \apply{\Pi}{I_{i-1}}}$. Then, ${\mat{\Pi}{E} =
\bigcup\textstyle_{i \geq 0} I_i}$ is the \emph{materialisation} of $\Pi$
w.r.t.\ $E$.

RDF \cite{rdf11-concepts} can represent graph-like data using \emph{triples}
${\langle s, p, o \rangle}$ where $s$, $p$, and $o$ are constants. Intuitively,
a triple says that a subject $s$ has a property $p$ with value $o$. An
\emph{RDF graph} is a finite set of triples. In this paper, we apply datalog to
RDF using \emph{vertical partitioning} \cite{DBLP:journals/vldb/AbadiMMH09}: we
convert each triple ${\langle s, p, o \rangle}$ to a unary fact $o(s)$ if ${p =
\mathit{rdf}{:}\mathit{type}}$, and otherwise to a binary fact $p(s,o)$. Due to
this close correspondence, we usually do not distinguish facts from triples.

\section{Our Approach}\label{sec:approach}

Our main idea is to represent the derived facts compactly using structure
sharing. Presenting the full details of our approach requires quite a bit of
notation, so we defer the presentation of all algorithms to
\iftoggle{withappendix}{Appendix~\ref{sec:formalisation}}{online appendix
\cite{compressed-extended}}, and in this section we present only the main ideas
on a running example. Assume we are given an RDF graph containing the following
triples.
\begin{multicols}{2}
\noindent
\begin{align*}
    \langle a_i, P, d \rangle                               & \text{ for } 1 \leq i \leq 2n \\
	\langle a_{2i}, \mathit{rdf}{:}\mathit{type}, R \rangle & \text{ for } 1 \leq i \leq n   
\end{align*}
\begin{align*}
	\langle b_i, P, c_i \rangle                             & \text{ for } 1 \leq i \leq m  \\
	\langle d, T, e_i \rangle                               & \text{ for } 1 \leq i \leq m  
\end{align*}
\end{multicols}
Using vertical partitioning described in Section~\ref{sec:preliminaries}, we
convert the above triples into a dataset $E$ containing explicit facts
\eqref{eq:in-1}--\eqref{eq:in-4}.
\begin{multicols}{2}
\noindent
\begin{align}
	P(a_i,d)    & \text{ for } 1 \leq i \leq 2n \label{eq:in-1} \\
	R(a_{2i})   & \text{ for } 1 \leq i \leq n  \label{eq:in-3}  
\end{align}
\begin{align}
	P(b_i,c_i)  & \text{ for } 1 \leq i \leq m  \label{eq:in-2} \\
	T(d, e_i)   & \text{ for } 1 \leq i \leq m  \label{eq:in-4} 
\end{align}
\end{multicols}
\noindent Finally, let $\Pi$ be a recursive program containing rules
\eqref{eq:rule-1} and \eqref{eq:rule-2}.
\begin{align}
	P(x,y) \wedge R(x)      & \rightarrow S(x,y) \label{eq:rule-1} \\
	S(x,y) \wedge T(y,z)    & \rightarrow P(x,z) \label{eq:rule-2}
\end{align}
Instead of computing $\mat{\Pi}{E}$ directly, we compute a compressed
representation of $E$, and then we compute the materialisation over this
representation to reduce the number of rule applications and the space required
to store the derived facts. We next describe the general framework, and then we
discuss key operations such as rule application and elimination of duplicate
facts. Note that both rules in $\Pi$ contain more than one body atom, so both
RDFox and VLog would evaluate the rules as usual.

\leanparagraph{Representation and Framework} All of our algorithms require an
arbitrary, but fixed total ordering $\prec$ over all constants. Typically, a
natural such ordering exists; for example, many RDF systems represent constants
by integer IDs, so $\prec$ can be obtained by comparing these IDs. In our
example, we assume that $a_1 \prec \dots \prec a_{2n} \prec b_1 \prec \dots
\prec b_m \prec c_1 \prec \dots \prec c_m \prec d \prec e_1 \prec \dots \prec
e_m$ holds.

Our compressed representation of facts draws inspiration from columnar
databases. For example, we represent facts \eqref{eq:in-2} using a single fact
${P(\mc b, \mc c)}$, where $\mc b$ represents a vector of constants ${b_1 \dots
b_m}$ and $\mc c$ represents ${c_1 \dots c_m}$. To distinguish ${P(\mc b, \mc
c)}$ from the facts it represents, we call the former a \emph{meta-fact}.
Meta-facts are constructed like ordinary facts, but they use
\emph{meta-constants} (e.g., $\mc b$ and $\mc c$), which represent vectors of
ordinary constants. We maintain a mapping $\mu$ of meta-constants to the
constants they represent; thus, we let ${\mu(\mc b) = b_1 \dots b_m}$ and
${\mu(\mc c) = c_1 \dots c_m}$.

This representation is thus far not inherently more efficient than storing each
fact separately: although we use just one meta-fact ${P(\mc b, \mc c)}$, we
must also store the mapping $\mu$ so the combined storage cost is the same.
However, this approach allows \emph{structure sharing}. For example, consider
applying rule ${P(x,y) \rightarrow W(x,y)}$ to our facts. A conventional
datalog system would derive $m$ new facts, whereas we can represent all
consequences of the rule by just one meta-fact ${W(\mc b,\mc c)}$ and thus
reduce the number of rule applications and the space needed. This case is
simple since the rule contains just one body atom. In this paper, we generalise
this idea to rules with several body atoms. To support efficient representation
and join computation, we introduce a richer way of mapping meta-constants to
constants. For a meta-constant $\mc a$, we allow $\mu(\mc a)$ to be (i)~a
vector of constants sorted by $\prec$, or (ii)~a vector of meta-constants.
Meta-constant $\mc a$ can thus be recursively \emph{unfolded} into a sorted
vector of constants that it represents. Since it is sorted by $\prec$, this
unfolding is unique. For example, if ${\mu(\mc a) = \mc g.\mc h}$, and
${\mu(\mc g) = a_1.a_3 \dots a_{2n-1}}$ and ${\mu(\mc h) = a_2.a_4 \dots
a_{2n}}$, then $a_1.a_2 \dots a_{2n}$ is the unfolding of $\mc a$. Moreover,
repeated constants can be stored using run-length encoding to reduce the space
requirements: we use $d * n$ to refer to constant $d$ repeated $n$ times.
Finally, we define the notion of meta-substitutions analogously to
substitutions, with a difference that variables are mapped to meta-constants
rather than constants.

Now we are ready to discuss how to generate a set of meta-facts $M$ and a
mapping $\mu$ for our example dataset $E$. For unary facts such as
\eqref{eq:in-3}, this is straightforward: we simply sort the facts by $\prec$,
we define $\mu(\mc h)$ as the vector of (sorted) constants $a2.a4 \dots
a_{2n}$, and we produce a meta-fact $R(\mc h)$. For binary facts, it is not
always possible to generate one meta-fact per predicate since one may not be
able to sort binary facts on both arguments simultaneously. For example,
sorting facts \eqref{eq:in-1} and \eqref{eq:in-2} on the first argument
produces a sequence $P(a_1,d) \dots P(a_{2n},d) \, P(b_1,c_1) \dots
P(b_m,c_m)$, which is not sorted on the second argument due to ${c_i \prec d}$.
Thus, we convert these facts into meta-facts by sorting the facts
lexicographically; we consider the argument with fewer distinct values first to
maximise the use of run-length encoding. Facts \eqref{eq:in-1}--\eqref{eq:in-2}
have fewer distinct values in the second argument, so we sort on that argument
first. Then, we iterate through the sorted facts and try to append each fact to
existing meta-facts, and we create fresh meta-facts when it is impossible to
find an appropriate such meta-fact. In our example, we thus obtain the following
meta-facts and mapping $\mu$.
\begin{align}
    P(\mc a, \mc d) & \qquad\quad P(\mc b, \mc c)   & R(\mc h)      & \qquad\quad T(\mc e, \mc f)   \\
	\mu(\mc a)      & = a_1.a_2 \dots a_{2n}        & \mu(\mc b)    & = b_1 \dots b_m               \\
    \mu(\mc c)      & = c_1 \dots c_m               & \mu(\mc d)    & = d * 2n                      \\
    \mu(\mc e)      & = d * m                       & \mu(\mc f)    & = e_1 \dots e_m               \\
	\mu(\mc h)      & = a_2.a_4 \dots a_{2n}
\end{align}

With this set of meta-facts $M$, mapping $\mu$, and program $\Pi$, we use a
variant of the semina\"ive algorithm~\cite{abiteboul95foundation} to compute
the materialisation over $M$---that is, we keep applying the rules of $\Pi$ to
$M$ until no further facts can be derived. To avoid applying a rule to a set of
facts more than once, we maintain a set $\Delta$ of meta-facts derived in the
previous round of rule application, and, when applying a rule, we require at
least one body atom to be matched to a meta-fact in $\Delta$. In each round of
rule application, we evaluate rule bodies as queries, where join evaluation is
accomplished using two new \emph{semi-join} and \emph{cross-join} algorithms.
Moreover, to correctly maintain $\Delta$, we apply duplicate elimination at the
end of each round. Note that this is critical for the termination of
materialisation: without duplicate elimination, a group of rules could
recursively derive the same facts and never terminate. We next run the above
process over our example and discuss each round of rule application in detail.

\leanparagraph{First Round} Set $M$ initially does not contain a meta-fact with
predicate $S$, so rule~\eqref{eq:rule-2} does not derive anything. To apply
rule~\eqref{eq:rule-1}, we note that all variables of atom $R(x)$ are contained
in the variables of atom $P(x,y)$, so we evaluate the rule body using a
semi-join, where $x$-values from $R(x)$ act as a filter on $P(x,y)$. We first
identify a set of substitutions that survive the join, and then we reorganise
the result so that it can be represented using structure sharing.

Matching atom $P(x,y)$ in rule~\eqref{eq:rule-1} produces meta-substitutions
$\sigma_1 = \{ x \mapsto \mc a, y \mapsto \mc d \}$ and $\sigma_2 = \{ x
\mapsto \mc b, y \mapsto \mc c \}$, and matching $R(x)$ produces $\sigma_3 = \{
x \mapsto \mc h\}$. Since the unfolding of each of the meta-constant is sorted
w.r.t. $\prec$, we can join these meta-substitutions using a merge-join. Thus,
we initialise a priority queue $F$ to contain the substitutions obtained from
$\sigma_1$ and $\sigma_2$ by replacing each meta-constant with the first
constant in the unfolding; thus, $F$ initially contains $\{x \mapsto a_1, y
\mapsto d\}$ and $\{x \mapsto b_1, y \mapsto c_1\}$. We analogously initialise
a priority queue $G$ with $\sigma_3$ to contain $\{x \mapsto a_2\}$. Our queues
$F$ and $G$ also record the meta-substitutions that the respective
substitutions come from. To perform the join, we iteratively select the
$\preceq_x$-least substitutions from $F$ and $G$ and compare them; if two
substitutions coincide on the common variables $x$, we add the substitution
from $F$ to the result set $S$; and we proceed to the next substitutions from
$F$ and/or $G$, as appropriate. After processing all of $F$ and $G$, set $S$
contains all substitutions that survive the join. In our running example, set
$S$ contains substitutions ${\{ x \mapsto a_{2i}, y \mapsto d \}}$ for ${1 \leq
i \leq n}$.

Thus, the $a_{2i}$ values in the unfolding of $\mc a$ have survived the join,
whereas the $a_{2i-1}$ values have not. To facilitate structure sharing, we
\emph{shuffle} meta-constant $\mc a$ by splitting it into two meta-constants
$\mc g$ and $\mc h$. We let $\mu(\mc h) = a_2.a_4 \dots a_{2n}$ to represent
the constants that have survived the join, and we let $\mu(\mc g) = a_1.a_3
\dots a_{2n-1}$ to represent the constants that have not survived. We redefine
the representation $\mc a$ by setting $\mu(\mc a) = \mc g. \mc h$; doing so
does not change the unfolding of $\mc a$. Finally, we introduce a new
meta-constant $\mc j$ and set $\mu(\mc j) = d * n$, so the rule conclusion can
be represented as $S(\mc h, \mc j)$. No meta-facts with $S$ predicate have been
derived to this point, so duplicate elimination is superfluous and we add
$S(\mc h, \mc j)$ to $\Delta$.

The above computation on our example requires $O(n)$ steps, which is the same
as in evaluating the rule on plain facts; however, the space requirement is
only $O(1)$ due to structure sharing, as opposed to $O(n)$ for the case of
normal join on plain facts.

\leanparagraph{Second Round} Set $\Delta$ does not contain $P$ or $R$
meta-facts, so rule~\eqref{eq:rule-1} is not matched in the second round.
However, in rule~\eqref{eq:rule-2}, we can match $S(x,y)$ to $S(\mc h, \mc j)$
in $\Delta$, and we can match $T(y,z)$ to $T(\mc e, \mc f)$. The two sets of
variables obtained from the two body atoms intersect, but neither of them
includes the other; thus, we evaluate the rule body by performing a cross-join.
As in the case of semi-join, we construct priority queues $F$ and $G$ to
iterate over all substitutions represented by meta-substitutions ${\{x \mapsto
\mc h, y \mapsto \mc j\}}$ and ${\{y \mapsto \mc e, z \mapsto \mc f\}}$,
respectively. Initially, $F$ contains $\{x \mapsto a_2, y \mapsto d\}$ and $G$
contains $\{y \mapsto d, z \mapsto e_1\}$. These two substitutions agree on
$y$, so we collect all substitutions represented by ${\{y \mapsto \mc e, z
\mapsto \mc f\}}$ where $y$ is mapped to $d$, and we compress the result. In
our example, all substitutions represented by $\{y \mapsto \mc e, z \mapsto \mc
f\}$ map $y$ to $d$. Then, we iterate through each substitution $\sigma$
represented by ${\{x \mapsto \mc h, y \mapsto \mc j\}}$ where $\sigma$ maps $y$
to $d$, and we produce a meta-substitution $\beta$ representing the join
between $\sigma$ and $\{z \mapsto \mc f\}$. We thus obtain $\{x \mapsto \mc
a_{2i}, y \mapsto \mc e, z \mapsto \mc f\}, 1 \leq i \leq n$, where $\mu(\mc
a_{2i}) = a_{2i}*m$, and so we represent the join result as $P(\mc a_{2i}, \mc
f)$.

Since the set of derived facts already contains $P(\mc a, \mc d)$ and $P(\mc b,
\mc c)$, to remove duplicates we compare the facts represented by the newly
derived $P$ meta-facts with those represented by the two existing meta-facts.
This is achieved by using priority queues to perform a merge-anti-join. On our
example no duplicates can be found, so we compute $\Delta$ as $\{P(\mc a_{2i},
\mc f), 1 \leq i \leq n\}$.

The above computation introduces $n$ new meta constants and $n$ new meta-facts,
and it requires only $O(n)$ space, as opposed to $O(n^2)$ needed to compute the
join over ordinary facts. Moreover, producing each new meta-fact takes only
$O(1)$ steps so our cross-join requires a total of $O(n)$ steps, instead of
$O(n^2)$. Finally, our duplicate elimination method still requires $O(n^2)$
time since the meta-facts must be unpacked and compared.

\leanparagraph{Termination} In the third round, we can match atom $P(x,y)$ to
$P(\mc a_{2i}, \mc f)$ in $\Delta$ and $R(x)$ to $R(\mc h)$, and derive $S(\mc
a_{2i}, \mc f)$ for $1 \leq i \leq n$. In the fourth round, we try to join
$S(\mc a_{2i}, \mc f)$ with $T(\mc e, \mc f)$, but nothing is derived, so the
fixpoint is reached. The derived meta-facts include $S(\mc h, \mc j)$, $P(\mc
a_{2i}, \mc f)$, and $S(\mc a_{2i}, \mc f)$, and $\mu$ is changed as follows.
\begin{align}
	\mu(\mc a)  & = \mc g. \mc h    & \mu(\mc g)        & = a_1.a_3 \dots a_{2n-1}                  \\
	\mu(\mc j)  & = d * n           & \mu(\mc a_{2i})   & = a_{2i} * m \text{ for } 1 \leq i \leq n
\end{align}
Our approach thus clearly only requires $O(n)$ space (rather than $O(n^2)$) for
storing the derived meta-facts and the the mapping $\mu$. Such saving can be
significant, particularly when $n$ is large.

\section{Evaluation}\label{sec:evaluation}

We have implemented our approach in a prototype system called CompMat and
compared it with VLog and RDFox, two most closely related state-of-the-art
systems. We considered two VLog variants: one stores the explicitly given facts
on disk in an RDF triple store, and another reads them from CSV files and
stores them in RAM; both VLog variants store the derived facts in RAM. RDFox is
purely RAM-based. Both systems use the semina\"ive algorithm.

\leanparagraph{Test Benchmarks} For our evaluation, we used benchmarks derived
from the following well-known RDF datasets.
LUBM~\cite{DBLP:journals/ws/GuoPH05} is a synthetic benchmark describing the
university domain. We used the 1K dataset. Reactome~\cite{croft2013reactome}
describes biochemical pathways of proteins, drugs, and other agents.
Claros~\cite{DBLP:conf/dihu/RahtzDKKZA11} is real-world dataset describing
cultural artefacts. We obtained the \emph{lower bound} (L) datalog programs
from the accompanying OWL ontologies as described by
\citet{motik2015incremental}---that is, we apply the sound but incomplete
transformation by \citet{GHVD03} without explicitly axiomatising the
\emph{owl:sameAs} relation. In addition, the Claros \emph{lower bound extended}
(LE) program was obtained by extending \emph{Claros\_L} with several
`difficult' rules. All our test programs are recursive.

\leanparagraph{Test Setup} For each benchmark and test system, we loaded the
dataset and the program into the system and computed the materialisation. For
each test run, we measured the wall-clock times for loading plus
materialisation. Both VLog and RDFox index the data during loading. In
contrast, CompMat does not perform any preprocessing during loading, and it
compresses the explicitly given facts as part of the materialisation process.

We also used a new \emph{representation size} metric to measure the compactness
of representation without taking into account any implementation-specific
issues such data indexing. This measure counts the symbols needed to encode the
facts. We can encode a dataset $I$ containing $n$ predicates, each of arity
$a_i$ and containing $m_i$ facts, as a `flat' list where we output each
predicate once and then list the arguments of all facts with that predicate;
thus, we define the representation size as ${\repsize{I} = \sum_{i=1}^n (1 +
a_i \cdot m_i)}$. Thus, $\repsize{\mat{\Pi}{E}}$ provides us with a baseline
measure. In our approach, $\mat{\Pi}{E}$ is represented as a pair ${\langle M,
\mu \rangle}$ of a set $M$ meta-facts and a mapping $\mu$. Since $M$ is a
dataset, we define $\repsize{M}$ as above. Moreover, we define $\repsize{\mu}$
as the sum of the sizes of the mappings for each meta-constant, each encoded
using run-length encoding. That is, if $\mu(\mc a)$ contains $m$ distinct
(meta-)constants, the representation size of the mapping for $\mc a$ is ${1 + 2
\cdot m}$ since we can encode the mapping as $\mc a$ followed by a sequence of
pairs of a (meta-)constant and the number of its repetitions. We use just one
symbol for the number of occurrences since both (meta-)constants and number of
occurrences are likely to be represented as fixed-width integers in practice.
To further analyse our approach, we also report the average length of the
unfolding of the meta-constants in $\mu$ after materialisation.

\setlength{\tabcolsep}{3pt}
\begin{table}[tb]
\centering
\small
\newcommand{\tr}[1]{\multirow{2}{*}{#1}}
\begin{tabular}{l|rrr|rrr|r}
    \tr{Dataset}    & $\repsize{E}$ & $\repsize{I}$ & Diff.     & $\repsize{\langle E, \mu \rangle}$    & $\repsize{\langle M, \mu \rangle}$    & Diff. & Avg.\         \\
                    & (M)           & (M)           & (M)       & (M)                                   & (M)                                   & (M)   & len.\ $\mu$   \\
    \hline
    LUBM-1K$_L$     & 241.3         &  314.4        &   70.3    &  195.2                                & 195.9                                 &   0.7 & 7992.8        \\
    Reactome$_L$    &  22.7         &   32.3        &    9.6    &   20.2                                &  25.1                                 &   4.9 &  21.9         \\
    Claros$_L$      &  32.2         &  105.5        &   73.3    &   28.1                                &  31.2                                 &   3.1 &  104.8        \\
    Claros$_{LE}$   &  32.2         & 1065.8        & 1033.6    &   28.1                                & 413.9                                 & 385.8 &   127.1       \\
    \hline
\end{tabular}
\caption{Dataset statistics: all numbers apart from the average length of $\mu$
are in millions.}\label{tab:dataset-sizes}
\vspace{-25pt}
\end{table}
\setlength{\tabcolsep}{6pt}
\begin{table}[tb]
\centering
\small
\begin{tabular}{l|r|r|r|r}
    Dataset         & CompMat           &    VLog (RDF)    &  VLog (CSV)       & RDFox      \\
    \hline
    LUBM-1K$_L$     & 266.8 \phantom{k} &    1233.7        &  300.1            &  488.3     \\
    Reactome$_L$    &  47.3 \phantom{k} &    44.0          &  27.5             &  53.0      \\
    Claros$_L$      &  59.1 \phantom{k} &    198.4         &  47.0            &  135.9     \\
    Claros$_{LE}$   &  10.2 k           &    2869.9        &  2684.0           &  3492.1    \\
    \hline
\end{tabular}
\caption{Performance of tested systems.}\label{tab:times}
\vspace{-25pt}
\end{table}

\leanparagraph{Test Results} Table~\ref{tab:dataset-sizes} shows the sizes of
the `flat' representation and our compact representation before and after
materialisation and their difference, as well as information about the mapping
$\mu$. Table~\ref{tab:times} shows the running times (in seconds) of all
systems.

As one can see, the representation size of the explicit facts is smaller in our
approach due to run-length encoding, but these savings are generally
negligible. In contrast, derived facts are represented much more compactly in
all cases: the 48.8~M derived facts in LUBM-1K$_L$ require just 0.7~M
additional symbols, instead of 70.3~M symbols needed for a `flat'
representation; 55~M derived facts in Claros$_L$ require just 3.1~M, instead of
73.3~M additional symbols; our technique uses about half as many additional
symbols on Reactome$_L$; and even on Claros$_{LE}$ it is very effective and
reduces the number of symbols by a factor of three. These results are reflected
in the structure of $\mu$: the average mapping length in above 100 on all
benchmarks apart from Reactome$_L$, which suggests a significant degree of
structure sharing.

In terms of the cumulative time, CompMat turned out to be fastest on
LUBM-1K$_L$ and very competitive on Claros$_L$. On Reacome$_L$ our system was
narrowly outperformed by VLog. In contrast, CompMat was considerably slower
than the other systems on Claros$_{LE}$. In all cases, we observed that our
system spends most of the time in duplicate elimination. Hence, it seems that
our representation can be very effective in reducing the number of rule
applications, but at the expense of more complex duplicate elimination.

\section{Conclusion}\label{sec:conclusion}

We have presented a new datalog materialisation technique that uses structure
sharing to represent derived facts. This not only allows for more compact
storage of facts, but also allows applying the rules without considering each
rule derivation separately. We have implemented our technique in a new system
called CompMat and have shown it to be competitive with VLog and RDFox. Also,
our representation was more compact than the `flat' representation in all
cases, sometimes by orders of magnitude.

\bibliographystyle{ACM-Reference-Format}
\bibliography{references}

\iftoggle{withappendix}{
    \newpage
    \appendix
    \section{Formalisation and Algorithms}\label{sec:formalisation}

We now formalise the ideas from Section~\ref{sec:approach}. Recall that a
meta-constant $\mc a$ can be recursively \emph{unfolded} into a sorted vector
of constants that it represents. Since it is sorted by $\prec$, this unfolding
is unique so we can identify a constant at some integer index in the unfolding
of $\mc a$. For example, if ${\mu(\mc a) = \mc g.\mc h}$, and ${\mu(\mc g) =
a_1.a_3 \dots a_{2n-1}}$ and ${\mu(\mc h) = a_2.a_4 \dots a_{2n}}$, then
$a_1.a_2 \dots a_{2n}$ is the unfolding of $\mc a$, and $a_3$ is the constant
at index $3$.

We next introduce several useful notions. Given a meta-constant $\mc a$, we let
$|\mc a|$ be the length of the unfolding of $\mc a$, and we let $\tail{\mc a}$
be the last constant in the unfolding. If $\mu(\mc a)$ is a sequence of
constants, then $\mc a$ is called a \emph{leaf meta-constant}. The
\emph{length} of a meta-fact is equal to the length of its meta-constants. A
\emph{meta-substitution} $\sigma$ is a mapping of variables to meta-constants
such that ${|\sigma(x)| = |\sigma(y)|}$ holds for all ${x,y \in \dom{\sigma}}$.
Moreover, ${|\sigma| = 0}$ if ${\dom{\sigma} = \emptyset}$, and otherwise
${|\sigma| = |\sigma(x)|}$ for some ${x \in \dom{\sigma}}$. Finally, for $B$ a
constant-free atom with no repeated variables and $M$ a set of (meta-)facts,
$\evaluate{B}{M}$ is the set of (meta-)substitutions $\sigma$ such that
${B\sigma \in M}$.

Based on these definitions, Algorithm~\ref{alg:cmat} accepts a program $\Pi$
and a set $E$ of explicitly given facts, and it computes a set $M$ of
meta-facts and a mapping $\mu$ that represent $\mat{\Pi}{E}$. To this end, we
first convert $E$ into meta-facts
(lines~\ref{alg:cmat:prepare:start}--\ref{alg:cmat:prepare:end}): for each
predicate $P$, we retrieve all substitutions corresponding to all $P$-facts in
$E$, we use function $\mathsf{compress}$ to convert them into one or more
meta-substitutions (line~\ref{alg:cmat:prepare:eval}), and we convert the
result back into meta-facts (line~\ref{alg:cmat:prepare:derive}). Compression
creates meta-constants by mapping constants to monotonically increasing
sequences: a substitution ${\sigma \in S}$ is appended to a meta-substitution
$\tau$ produced thus far (line~\ref{alg:compress:append}) if, for each ${x \in
\dom{\sigma}}$, constant $\sigma(x)$ is larger than or equal to the last
constant in the sequence $\mu(\tau(x))$ (line~\ref{alg:compress:append-check});
otherwise, we create a fresh meta-substitution to represent $\sigma$
(line~\ref{alg:compress:new:tau}).

We next apply the rules of $\Pi$ up to the fixpoint
(line~\ref{alg:cmat:apply:start}--\ref{alg:cmat:apply:end}). We use a variant
of the well-known \emph{semina\"ive} algorithm \cite{abiteboul95foundation} to
avoid redundant rule applications: we maintain a set $\Delta$ of meta-facts
derived in the previous round of rule application, and in each round we require
each rule to match at least one body atom in $\Delta$. To this end, we consider
rule ${r \in \Pi}$ and each atom $B_i \in \body{r}$
(lines~\ref{alg:cmat:rule:start}--\ref{alg:cmat:rule:end}), and we evaluate
$\body{r}$ left-to-right by matching each atom $B_j$ before $B_i$ in ${M
\setminus \Delta}$, atom $B_i$ in $\Delta$, and each atom $B_j$ after $B_i$ in
$M$ (lines~\ref{alg:cmat:atom:start}--\ref{alg:cmat:atom:end}). We discuss the
function $\mathsf{match}$ in Section~\ref{sec:sjoin}. During this process, set
$L$ keeps the meta-substitutions corresponding to the matches of atoms up to
$B_j$. Moreover, set $V$ keeps the variables matched thus far. We use $V$ to
decide how to join atom $B_j$ with $L$: we use a \emph{semi-join} if the
variables of one side of the join are contained in the variables of the other
side (lines~\ref{alg:cmat:sjoin1} and~\ref{alg:cmat:sjoin2}), and otherwise we
use a more general \emph{cross-join} (line~\ref{alg:cmat:xjoin}). These
algorithms are two main novel aspects of our approach, and we describe them in
detail in Sections~\ref{sec:sjoin} and~\ref{sec:xjoin}. After processing all
body atoms of $r$, we convert set $L$ into meta-facts corresponding to the head
of $r$ (line~\ref{alg:cmat:rule:N}). After applying all rules, newly derived
meta-facts are subjected to duplicate elimination
(line~\ref{alg:cmat:elimDup}), which we describe in
Section~\ref{sec:duplicates}. Finally, all meta-facts over meta-constants of
length one are removed from $M$, compressed using Algorithm~\ref{alg:compress},
and added back to $M$ (line~\ref{alg:cmat:recompress}). This step turned out to
be critical to the performance of our approach by reducing the number of
meta-facts in $M$, which in turn improved the speed of join computation.

\begin{figure}[tb]
\vspace*{-\baselineskip}
\begin{algorithm}[H]
\caption{$\cmat{\Pi, E}$}\label{alg:cmat}
\begin{algorithmiccont}[1]
    \State $M \defeq \emptyset$, \quad $\mu \defeq \emptyset$                                                   \label{alg:cmat:prepare:start}
    \For{\textbf{each} $n$-ary predicate $P$ occurring in $E$}
        \State $A \defeq P(x_1, \dots, x_n)$
        \For{\textbf{each} $\tau \in \compress{\evaluate{A}{E}, \mu}$}                                          \label{alg:cmat:prepare:eval}
            $M \defeq M \cup \{ A\tau \}$                                                                       \label{alg:cmat:prepare:derive}
        \EndFor
    \EndFor                                                                                                     \label{alg:cmat:prepare:end}
    \State $\Delta \defeq M$
    \While{$\Delta \neq \emptyset$}                                                                             \label{alg:cmat:apply:start}
        \State $N \defeq \emptyset$
        \For{\textbf{each} rule $B_1 \wedge \dots \wedge B_n \rightarrow H \in \Pi$ and ${1 \leq i \leq n}$}    \label{alg:cmat:rule:start}
            \State $L \defeq \{ \sigma_0 \}$ where $\sigma_0$ is the empty meta-substitution
            \State $V \defeq \emptyset$
            \For{\textbf{each} $1 \leq j \leq n$}
                \If{$j < i$}                                                                                    \label{alg:cmat:atom:start}
                   \hspace{0.45cm} $R \defeq \match{B_j}{M \setminus \Delta}$
                \ElsIf{$j = i$}
                   $R \defeq \match{B_j}{\Delta}$
                \Else
                   \hspace{1.6cm} $R \defeq \match{B_j}{M}$
                \EndIf                                                                                          \label{alg:cmat:atom:end}
                \If{$V \defeq \emptyset$}
                    \hspace{0.9cm} $L \defeq R$
                \ElsIf{$V \subseteq \vars{B_j}$}
                    $L \defeq \sjoin{L,R,V,M,\mu}$                                                              \label{alg:cmat:sjoin1}
                \ElsIf{$\vars{B_j} \subseteq V$}
                    $L \defeq \sjoin{R,L,\vars{B_j},M,\mu}$                                                     \label{alg:cmat:sjoin2}
                \Else
                    \hspace{2.25cm} $L \defeq \xjoin{L,R,V \cap \vars{B_j}, \mu}$                               \label{alg:cmat:xjoin}
                \EndIf
                \State $V \defeq V \cup \vars{B_j}$
            \EndFor
            \State $N \defeq N \cup \{ H\sigma \mid \sigma \in L \}$                                            \label{alg:cmat:rule:N}
        \EndFor                                                                                                 \label{alg:cmat:rule:end}
        \State $\Delta \defeq \Call{ElimDup}{N, M, \mu}$                                                        \label{alg:cmat:elimDup}
        \State $M \defeq M \cup \Delta$
        \State Compress all meta-facts in $M$ of length one                                                     \label{alg:cmat:recompress}
    \EndWhile                                                                                                   \label{alg:cmat:apply:end}
\end{algorithmiccont}
\end{algorithm}
\vspace{-0.9cm}
\begin{algorithm}[H]
\caption{$\compress{S,\mu}$}\label{alg:compress}
\begin{algorithmiccont}[1]
    \State $T \defeq \emptyset$
    \For{\textbf{each} substitution $\sigma \in S$}
        \If{there exists a meta-substitution $\tau \in T$ such that                                             \label{alg:compress:append-check}
                \Statex \hspace{1cm} $\tail{\tau(x)} \preceq \sigma(x)$ holds for each $x \in \dom{\sigma}$}
            \For{\textbf{each} $x \in \dom{\sigma}$}
                Append $\sigma(x)$ to  $\mu(\tau(x))$                                                           \label{alg:compress:append}
            \EndFor
        \Else
            \State Let $\tau$ be a meta-substitution where, for $x \in \dom{\sigma}$,                           \label{alg:compress:new:tau}
            \Statex \hspace{1cm}$\tau(x)$ is a fresh meta-constant and let $\mu(\tau(x)) \defeq \sigma(x)$
            \State $T \defeq T \cup \{ \tau \}$
        \EndIf
    \EndFor
    \State \Return $T$
\end{algorithmiccont}
\end{algorithm}
\vspace{-0.6cm}
\end{figure}

\subsection{Computing Semi-Joins}\label{sec:sjoin}

Function $\mathsf{sjoin}$ from Algorithm~\ref{alg:cmat} computes the semi-join
of sets $L$ and $R$ of meta-substitutions, where ${\dom{\lambda} \subseteq
\dom{\rho}}$ holds for all meta-substitutions ${\lambda \in L}$ and ${\rho \in
R}$; the vector $\vec x$ contains all variables common to the substitutions in
$L$ and $R$. Set $L$ thus acts as a filter on $R$: we identify a set $S$ of
substitutions represented by $R$ that survive the join, and we reorganise the
representation so that the result can be represented using structure sharing.
We need additional notation to formalise this idea.

Please remember that $\prec$ is the ordering on constants from
Section~\ref{sec:formalisation}. Then, for ${\vec x = x_1, \dots, x_n}$ a
vector of variables, we define an ordering $\prec_{\vec x}$ on substitutions
such that ${\xi \prec_{\vec x} \zeta}$ holds for substitutions $\xi$ and
$\zeta$ iff there exists ${1 \leq i \leq n}$ such that ${\xi(x_j) =
\zeta(x_j)}$ for each ${1 \leq j < i}$ and ${\xi(x_i) \prec \zeta(x_i)}$. That
is, $\prec_{\vec x}$ compares substitutions lexicographically by $\vec x$.
Analogously, ${\xi =_{\vec x} \zeta}$ holds iff ${\xi(x_i) = \zeta(x_i)}$ for
${1 \leq i \leq n}$.

For $\sigma$ a meta-substitution and $i$ an integer, we define
$\subst{\sigma}{i}$ as the $i$-th substitution that $\sigma$ represents---that
is, for ${x \in \dom{\sigma}}$, each $\subst{\sigma}{i}(x)$ is the $i$-th
constant from the unfolding of $\mu(\sigma(x))$.

Finally, we use priority queues of pairs of the form ${\langle \sigma, i
\rangle}$ where $\sigma$ is a meta-substitution and ${1 \leq i \leq |\sigma|}$.
Such ${\langle \sigma, i \rangle}$ represents $\subst{\sigma}{i}$, but it
maintains the separation of $\sigma$ and $i$ so we can enumerate the
substitutions that $\sigma$ represents. For $\vec x$ a vector of variables, let
${\langle \sigma, i \rangle \prec_{\vec x} \langle \tau, j \rangle}$ iff
${\subst{\sigma}{i} \prec_{\vec x} \subst{\tau}{j}}$. Given a set $S$ of such
pairs, $\queue{\vec x}{S}$ creates a queue $Q$ that contains $S$ sorted by
$\prec_{\vec x}$. Moreover, $\peekQ{Q}$ returns a $\preceq_{\vec x}$-smallest
pair ${\langle \sigma, i \rangle \in Q}$; if there are several such pairs
(which is possible if $\vec x$ does not cover all variables of $\sigma$), then
one arbitrarily chosen, but fixed pair is returned. Finally, $\nextQ{Q}$
removes this pair ${\langle \sigma, i \rangle}$ from $Q$, adds ${\langle
\sigma, i+1 \rangle}$ to $Q$ if ${i+1 \leq |\sigma|}$, reorders $Q$ so it is
sorted by $\preceq_{\vec x}$, and returns ${\langle \sigma, i \rangle}$.

Algorithm~\ref{alg:sjoin} computes the semi-join of $L$ and $R$. Since
lines~\ref{alg:cmat:sjoin1} and~\ref{alg:cmat:sjoin2} pass a set of variables
$V$ for $\vec x$, to bridge this gap we assume that the variables of $V$ are
ordered in some way when calling $\mathsf{sjoin}$. To compute the semi-join, we
initialise priority queues $F$ and $G$ to contain the first substitutions
represented by the meta-substitutions in $L$ and $R$, respectively
(lines~\ref{alg:sjoin:init:F}--\ref{alg:sjoin:init:G}). Now, meta-constants are
mapped to increasing sequences of constants w.r.t.\ $\preceq$, so
${\subst{\sigma}{i} \preceq_{\vec x} \subst{\sigma}{j}}$ holds for each $\vec
x$, $\sigma$, and ${i \leq j}$. Thus, we can join $F$ and $G$ using merge-join
(lines~\ref{alg:sjoin:loop:start}--\ref{alg:sjoin:loop:end}): we select the
$\preceq_{\vec x}$-least pairs ${\langle \lambda,i \rangle}$ and ${\langle
\rho,j \rangle}$ of $F$ and $G$ (line~\ref{alg:sjoin:least-pairs}) and compare
them; we add ${\langle \lambda,i \rangle}$ to $S$ if $\subst{\lambda}{i}$ and
$\subst{\rho}{j}$ coincide on the common variables $\vec x$
(line~\ref{alg:sjoin:add-S}); and we move to the next pair from $F$ and/or $G$,
as appropriate. After processing $F$ and $G$, set $S$ contains all
substitutions that survive the join.

Algorithm~\ref{alg:shuffle} converts $S$ into meta-substitutions with structure
sharing. For each meta-substitution $\rho$ in $S$, we compute the set $X$ of
indexes of substitutions represented by $\rho$ that `survive' the join
(line~\ref{alg:shuffle:rho-X}). We return $\rho$ if all substitutions `survive'
(line~\ref{alg:shuffle:sigma:rho}). Otherwise, for each variable ${x \in
\dom{\rho}}$ (line~\ref{alg:shuffle:x}), we unfold $\mu(\rho(x))$ and consider
each leaf meta-constant $\mc a_i$ encountered (line~\ref{alg:shuffle:a-i}). We
split $\mc a_i$ using two fresh meta-constants $\inmc{\mc b_i}$ and $\outmc{\mc
b_i}$ (lines~\ref{alg:shuffle:split:start}--\ref{alg:shuffle:split:end}): we
define $\mu(\inmc{\mc b_i})$ as the constants of $\mu(\mc a_i)$ at positions in
$X$ (i.e., the positions that survive the join), we define $\mu(\outmc{\mc
b_i})$ as the remaining constants of $\mu(\mc a_i)$, and we redefine $\mu(\mc
a_i)$ as $\inmc{\mc b_i}.\outmc{\mc b_i}$. This keeps the unfolding of $\mu(\mc
a_i)$ and of $\rho(x)$ unchanged, but it allows us to define the resulting
meta-substitution $\sigma$ on $x$ as the concatenation of all $\inmc{\mc b_i}$
(line~\ref{alg:shuffle:sigma:define}). Note that we can take $\inmc{\mc b_1}$
instead of introducing $\mc c$ whenever ${n = 1}$ holds.

\begin{figure}[tb]
\vspace*{-\baselineskip}
\begin{algorithm}[H]
\caption{$\sjoin{L, R, \vec x, M, \mu}$}\label{alg:sjoin}
\begin{algorithmiccont}[1]
    \State $F \defeq \queue{\vec x}{\{ \langle \lambda,1 \rangle \mid \lambda \in L \}}$                                            \label{alg:sjoin:init:F}
    \State $G \defeq \queue{\vec x}{\{ \langle \rho,1 \rangle \mid \rho \in  R \}}$                                                 \label{alg:sjoin:init:G}
    \State $S \defeq \emptyset$
    \While{$F \neq \emptyset$ and $G \neq \emptyset$}                                                                               \label{alg:sjoin:loop:start}
        \State $\langle \lambda,i \rangle \defeq \peekQ{F}$ and $\langle \rho,j \rangle \defeq \peekQ{G}$                             \label{alg:sjoin:least-pairs}
        \If{$\subst{\lambda}{i} \prec_{\vec x} \subst{\rho}{j}$}
            $\nextQ{F}$
        \Else
            \If{$\subst{\lambda}{i} =_{\vec x} \subst{\rho}{j}$}
                Add $\langle \rho,j \rangle$ to S                                                                                   \label{alg:sjoin:add-S}
            \EndIf
            \State $\nextQ{G}$
        \EndIf
    \EndWhile                                                                                                                       \label{alg:sjoin:loop:end}
    \State \Return $\shuffle{S, M, \mu}$
\end{algorithmiccont}
\end{algorithm}
\vspace{-0.9cm}
\begin{algorithm}[H]
\caption{$\shuffle{S, M, \mu}$}\label{alg:shuffle}
\begin{algorithmiccont}[1]
    \State $T \defeq \emptyset$
    \For{\textbf{each} distinct $\rho$ in $S$ and $X \defeq \{ j \mid \langle \rho, j \rangle \in S \}$}                            \label{alg:shuffle:rho-X}
        \If{$X = \{ 1, \dots, |\rho| \}$}                                                                                           \label{alg:shuffle:sigma:rho}
            Add $\rho$ to $T$
        \Else
            \State $\sigma \defeq \emptyset$
            \For{\textbf{each} variable $x \in \dom{\rho}$}                                                                         \label{alg:shuffle:x}
                \For{\textbf{each} leaf meta-constant $\mc a_i$ in $\mu(\rho(x))$}                                                  \label{alg:shuffle:a-i}
                    \State Introduce fresh meta-constants $\inmc{\mc b_i}$ and $\outmc{\mc b_i}$                                    \label{alg:shuffle:split:start}
                    \State Define $\mu(\inmc{\mc b_i})$ (resp.\ $\mu(\outmc{\mc b_i})$) as the sorted sequence
                    \Statex \hspace{1.4cm} of constants of $\mu(\mc a_i)$ whose corresponding indexes
                    \Statex \hspace{1.4cm} in $\mu(\rho(x))$ are contained (resp.\ not contained) in $X$
                    \State Redefine $\mu$ on $\mc a_i$ as $\mu(\mc a_i) \defeq \inmc{\mc b_i}.\outmc{\mc b_i}$                      \label{alg:shuffle:split:end}
                \EndFor
                \State Introduce a fresh meta-constant $\mc c$, define $\mu$ on $\mc c$ as
                \Statex \hspace{2.5cm} $\mu(\mc c) \defeq \inmc{\mc b_1}, \dots, \inmc{\mc b_n}$, and let $\sigma(x) \defeq \mc c$  \label{alg:shuffle:sigma:define}
            \EndFor
            \State Add $\sigma$ to $T$
        \EndIf
    \EndFor
    \State \Return $T$
\end{algorithmiccont}
\end{algorithm}
\vspace{-0.6cm}
\end{figure}

We finally discuss function $\match{B}{M}$ from
lines~\ref{alg:cmat:atom:start}--\ref{alg:cmat:atom:end} of
Algorithm~\ref{alg:cmat}. If atom $B$ has no repeated variables, we just return
$\evaluate{B}{M}$. Otherwise, we let $B'$ be an atom obtained by from $B$ by
renaming apart the repeated variables; we compute ${R \defeq
\evaluate{B'}{M}}$; we identify the set $S$ of pairs ${\langle \rho, i
\rangle}$ where ${\rho \in R}$ and $\subst{\rho}{i}$ satisfies variable
repetition; and we return $\shuffle{S, M, \mu}$. In other words, we reshuffle
the meta-facts that match $B$ so we can represent the matching portion using
structure sharing.

\subsection{Computing Cross-Joins}\label{sec:xjoin}

Function $\mathsf{xjoin}$ is used in Algorithm~\ref{alg:cmat} to compute the
cross-join of sets $L$ and $R$ of meta-substitutions with common variables
$\vec x$. We group the substitutions represented by the meta-substitutions in
$R$ on $\vec x$ and compress them as in Section~\ref{sec:formalisation}; this
allows us to avoid repetitions in the representation when computing the join
with the substitutions represented by the meta-substitutions in $L$.

This is captured in Algorithm~\ref{alg:xjoin}. As in Algorithm~\ref{alg:sjoin},
we construct priority queues $F$ and $G$
(lines~\ref{alg:xjoin:F}--\ref{alg:xjoin:G}) to iterate over all substitutions
represented by $L$ and $R$. We then use a variant of merge-join: we iteratively
select $\preceq_{\vec x}$-least pairs ${\langle \lambda,i \rangle}$ and
${\langle \rho,j \rangle}$ from $F$ and $G$ (line~\ref{alg:xjoin:least-pairs}),
and we advance $F$ or $G$ as needed if $\subst{\lambda}{i}$ and
$\subst{\rho}{j}$ do not agree on $\vec x$
(lines~\ref{alg:xjoin:advance:start}--\ref{alg:xjoin:advance:end}). Otherwise,
we collect all ${\langle \beta, k \rangle \in G}$ such that $\subst{\beta}{k}$
is equal to $\subst{\rho}{j}$ on $\vec x$ and remove the join variables
(lines~\ref{alg:xjoin:collect-R:start}--\ref{alg:xjoin:collect-R:end}), and we
compress the result (line~\ref{alg:xjoin:compress}) using
Algorithm~\ref{alg:compress}. We finally consider each ${\langle \alpha, \ell
\rangle \in F}$ such that $\subst{\alpha}{\ell}$ agrees with
$\subst{\lambda}{i}$ on $\vec x$
(lines~\ref{alg:xjoin:L:start}--\ref{alg:xjoin:L:end}) and, for each compressed
meta-substitution $\beta$, we produce a meta-substitution $\sigma$ representing
the join between $\subst{\lambda}{i}$ and all substitutions represented by
$\beta$ (lines~\ref{alg:xjoin:beta:start}--\ref{alg:xjoin:beta:end}).

\begin{algorithm}[tb]
\caption{$\xjoin{L, R, \vec x, \mu}$}\label{alg:xjoin}
\begin{algorithmiccont}[1]
    \State $F \defeq \queue{\vec x}{\{ \langle \lambda,1 \rangle \mid \lambda \in L \}}$                                                    \label{alg:xjoin:F}
    \State $G \defeq \queue{\vec x}{\{ \langle \rho,1 \rangle \mid \rho \in R \}}$                                                          \label{alg:xjoin:G}
    \State $S \defeq \emptyset$
    \While{$F \neq \emptyset$ and $G \neq \emptyset$}
        \State $\langle \lambda,i \rangle \defeq \peekQ{F}$ and $\langle \rho,j \rangle \defeq \peekQ{G}$                                     \label{alg:xjoin:least-pairs}
        \If{$\subst{\lambda}{i} \prec_{\vec x} \subst{\rho}{j}$}                                                                            \label{alg:xjoin:advance:start}
            $\nextQ{F}$
        \ElsIf{$\subst{\rho}{j} \prec_{\vec x} \subst{\lambda}{i}$}
            $\nextQ{G}$                                                                                                                      \label{alg:xjoin:advance:end}
        \Else
            \State $T \defeq \emptyset$
            \While{$\subst{\beta}{k} =_{\vec x} \subst{\rho}{j}$ for $\langle \beta, k \rangle \defeq \nextQ{G}$}                            \label{alg:xjoin:collect-R:start}
                \State Add $\subst{\beta}{k}$ restricted to the variables not in $\vec x$ to $T$
            \EndWhile                                                                                                                       \label{alg:xjoin:collect-R:end}
            \State $C \defeq \compress{T,\mu}$                                                                                              \label{alg:xjoin:compress}
            \While{$\subst{\alpha}{\ell} =_{\vec x} \subst{\lambda}{i}$ for $\langle \alpha,\ell \rangle \defeq \nextQ{F}$}                  \label{alg:xjoin:L:start}
                \For{\textbf{each} $\beta \in C$}                                                                                           \label{alg:xjoin:beta:start}
                    \State $\sigma \defeq \beta$
                    \For{\textbf{each} $x \in \dom{\lambda}$}
                        \State Introduce a fresh meta-constant $\mc a_x$, define $\mu(\mc a_x)$
                        \Statex \hspace{1.7cm} as $\subst{\alpha}{\ell}(x)$ repeated $|\beta|$ times, and let $\sigma(x) \defeq \mc a_x$
                    \EndFor
                    \State Add $\sigma$ to $S$
                \EndFor                                                                                                                     \label{alg:xjoin:beta:end}
            \EndWhile                                                                                                                       \label{alg:xjoin:L:end}
        \EndIf
    \EndWhile
    \State \Return $S$
\end{algorithmiccont}
\end{algorithm}

\subsection{Eliminating Duplicate Facts} \label{sec:duplicates}

Algorithm~\ref{alg:elimDup} is the final component of our approach: it takes
sets of meta-facts $N$ and $M$, and it returns the set $\Delta$ of meta-facts
representing all facts that are represented by $N$, but not by $M$. This is
critical for termination: datalog rules can be recursive, so facts produced by
a rule can (directly or indirectly) trigger further derivations using the same
rule; thus, if duplicate facts were not eliminated, a group of rules could keep
deriving the same facts indefinitely.

To this end, we consider each predicate $P$ in $N$ (line~\ref{alg:elimDup:P}),
and we eliminate all duplicate $P$-facts by perform a merge-anti-join between
$N$ and $M$ analogously to Algorithm~\ref{alg:sjoin}. In particular, we
initialise queues $F$ and $G$ so we can iterate over all facts represented by
$N$ and $M$ (lines~\ref{alg:elimDup:F}--\ref{alg:elimDup:G}), and we enumerate
the facts in $N$ by considering the corresponding ${\langle \lambda,i \rangle
\in F}$
(lines~\ref{alg:elimDup:enumerate-F:start}--\ref{alg:elimDup:enumerate-F:end}).
If $G$ is not empty, we skip all facts in $G$ that precede $\subst{\lambda}{i}$
in $\prec_{\vec x}$ (line~\ref{alg:elimDup:skip-G:start}), and we add ${\langle
\lambda, i \rangle}$ to $S$ if we find do not find a matching fact in $G$
(line~\ref{alg:elimDup:dup-detected} and~\ref{alg:elimDup:add-to-S}). Finally,
set $N$ can itself contain duplicate facts, so we skip all of those that match
$\subst{\lambda}{i}$ (line~\ref{alg:elimDup:skip-F:start}). After all facts in
$F$ have been considered, $S$ represents all distinct facts from $N$, so we use
shuffling from Section~\ref{sec:sjoin} to efficiently represent the result.

\begin{algorithm}[tb]
\caption{$\elimdup{N,M,\mu}$}\label{alg:elimDup}
\begin{algorithmiccont}[1]
    \State $\Delta \defeq \emptyset$
    \For{\textbf{each} $n$-ary predicate $P$ occurring in $N$}                                                  \label{alg:elimDup:P}
        \State Let $\vec x \defeq x_1. \dots. x_n$ be a vector of $n$ distinct variables
        \State $A \defeq P(\vec x)$, \qquad $S \defeq \emptyset$
        \State $F \defeq \queue{\vec x}{\{ \langle \lambda, 1 \rangle \mid \lambda \in \evaluate{A}{N} \}}$     \label{alg:elimDup:F}
        \State $G \defeq \queue{\vec x}{\{ \langle \rho, 1 \rangle \mid \rho \in \evaluate{A}{M} \}}$           \label{alg:elimDup:G}
        \While{$F \neq \emptyset$}                                                                              \label{alg:elimDup:enumerate-F:start}
            \State $\langle \lambda,i \rangle \defeq \peekQ{F}$
            \State $\mathit{notDup} \defeq \true$
            \If{$G \neq \emptyset$}
                \While{$\peekQ{G} \prec_{\vec x} \subst{\lambda}{i}$}                                            \label{alg:elimDup:skip-G:start}
                    $\nextQ{G}$
                \EndWhile                                                                                       \label{alg:elimDup:skip-G:end}
                \If{$\peekQ{G} =_{\vec x} \subst{\lambda}{i}$}
                    $\mathit{notDup} \defeq \false$                                                             \label{alg:elimDup:dup-detected}
                \EndIf
            \EndIf
            \If{$\mathit{notDup}$}
                Add $\langle \lambda,i \rangle$ to $S$                                                          \label{alg:elimDup:add-to-S}
            \EndIf
            \While{$\subst{\lambda}{i} =_{\vec x} \peekQ{F}$}                                                    \label{alg:elimDup:skip-F:start}
                $\nextQ{F}$
            \EndWhile                                                                                           \label{alg:elimDup:skip-F:end}
        \EndWhile                                                                                               \label{alg:elimDup:enumerate-F:end}
        \For{$\sigma \in \Call{Shuffle}{S, M, \mu}$}
            Add $A\sigma$ to $\Delta$
        \EndFor
    \EndFor
    \State \Return $\Delta$
\end{algorithmiccont}
\end{algorithm}

\section{Full Evaluation Results} \label{sec:full-results}

\begin{table*}[tb]
\centering
\small
\newcommand{\tr}[1]{\multirow{2}{*}{#1}}
\begin{tabular}{l|rrr|rrr|rrr|rrr}
    \tr{Dataset}    & $|\Pi|$   & $|E|$ & $|I|$ & $\repsize{E}$ & $\repsize{I}$ & Diff.     & $\repsize{\langle E, \mu \rangle}$    & $\repsize{\langle M, \mu \rangle}$    & Diff. & Avg.\         & Max.\         & Max.\ \\
                    &           & (M)   & (M)   & (M)           & (M)           & (M)       & (M)                                   & (M)                                   & (M)   & len.\ $\mu$   & len.\ $\mu$   & depth $ \mu$ \\
    \hline
    LUBM-1K$_L$     &   98      & 133.6 & 182.4 & 241.3         &  314.4        &   70.3    &  195.2                                & 195.9                                 &   0.7 & 7992.8        &  11.2 M       & 3    \\
    Reactome$_L$    &  541      &  12.5 &  19.8 &  22.7         &   32.3        &    9.6    &   20.2                                &  25.1                                 &   4.9 &   21.9        & 703.5 \;k     & 54   \\
    Claros$_L$      & 1310      &  18.8 &  73.8 &  32.2         &  105.5        &   73.3    &   28.1                                &  31.2                                 &   3.1 &  104.8        & 699.0 \;k     & 96   \\
    Claros$_{LE}$   & 1337      &  18.8 & 533.3 &  32.2         & 1065.8        & 1033.6    &   28.1                                & 413.9                                 & 385.8 &  127.1        & 699.0 \;k     & 2268 \\
    \hline
\end{tabular}
\caption{Dataset statistics: $|\Pi|$ is the number of rules; $I = \mat{\Pi}{E}$
is the materialised set of facts; and $|E|$ and $|I|$ are the numbers of facts
before and after materialisation. All numbers apart from $|\Pi|$ and the
statistics about $\mu$ are in millions.}\label{tab:dataset-sizes-extended}
\end{table*}

\begin{table*}[tb]
\centering
\small
\begin{tabular}{l|rrr|rrr|rrr|rrr}
    Dataset         & \multicolumn{3}{c|}{CompMat}                      & \multicolumn{3}{c|}{VLog (RDF)}       & \multicolumn{3}{c|}{VLog (CSV)}       & \multicolumn{3}{c}{RDFox} \\
                    & $t_l$     & $t_m$ \phantom{k} & $t_l + t_m$       & $t_l$     & $t_m$     & $t_l + t_m$   & $t_l$     & $t_m$     & $t_l + t_m$   & $t_l$     & $t_m$     & $t_l + t_m$   \\
    \hline
    LUBM-1K$_L$     & 198.0     & 68.8 \phantom{k}  & 266.8 \phantom{k} & 1211.0    &   22.7    & 1233.7        & 265.0     &   35.1    &  300.1        & 355.0     &  133.3    &  488.3        \\
    Reactome$_L$    & 20.3      & 27.0 \phantom{k}  &  47.3 \phantom{k} &   39.2    &    4.8    &   44.0        &  21.0     &    6.5    &   27.5        &  33.5     &   19.5    &   53.0        \\
    Claros$_L$      & 26.8      & 32.3 \phantom{k}  &  59.1 \phantom{k} &  189.7    &    8.7    &  198.4        &  33.0     &   14.0    &   47.0        &  47.1     &   88.8    &  135.9        \\
    Claros$_{LE}$   & 26.8      & 10.2 k            &  10.2 k           &  189.7    & 2680.2    & 2869.9        &  34.0     & 2650.0    & 2684.0        &  47.1     & 3445.0    & 3492.1        \\
    \hline
\end{tabular}
\caption{Performance of tested systems: $t_l$ and $t_m$ are loading and
materialisation times in seconds.}\label{tab:times-extended}
\end{table*}

We conducted our experiments on a Dell PowerEdge R720 server with 256~GB of RAM
and two Intel Xeon E5-2670 2.6 GHz processors, running Fedora 27 with kernel
version 4.15.12-301.fc27.x86\_64.
 
Table~\ref{tab:dataset-sizes-extended} extends Table~\ref{tab:dataset-sizes}
with statistics about our datasets. In particular, in
Table~\ref{tab:dataset-sizes-extended} we also report the maximum length of the
unfolding of the meta-constants in $\mu$, as well as the maximum meta-constant
depth: the depth of $\mc a$ is one if $\mc a$ is a leaf meta-constants, and the
depth of ${\mu(\mc a) = \mc b_1.\dots.\mc b_n}$ is one plus the maximum of the
depth of each $\mc b_i$.

Table~\ref{tab:times-extended} extends Table~\ref{tab:times} by showing
separately the loading ($t_l$) and materialisation ($t_m$) times. Please note
that both VLog and RDFox index the data during loading. In contrast, CompMat
does not perform any preprocessing during loading, and it compresses the
explicitly given facts as part of the materialisation process.

}{}

\end{document}